\documentclass[12pt]{article}

\usepackage{amsmath,amssymb}
\usepackage{graphicx}

\makeatletter

\renewcommand\section{\@startsection{section}{1}{\z@}
                                   {-3.5ex \@plus -1ex \@minus -.2ex}
                                   {2.3ex \@plus .2ex}
                                   {\normalfont\large\bfseries}}
\renewcommand\subsection{\@startsection{subsection}{2}{\z@}
                                   {-3.25ex\@plus -1ex \@minus -.2ex}
                                   {1.5ex \@plus .2ex}
                                   {\normalfont\normalsize\bfseries}}
\renewcommand\subsubsection{\@startsection{subsubsection}{3}{\z@}
                                   {-3.25ex\@plus -1ex \@minus -.2ex}
                                   {1.5ex \@plus .2ex}
                                   {\normalfont\normalsize\bfseries}}
\renewcommand\paragraph{\@startsection{paragraph}{4}{\z@}
                                   {3.25ex \@plus1ex \@minus.2ex}
                                   {-1em}
                                   {\normalfont\normalsize\bfseries}}

\makeatother

\newdimen\tableauside\tableauside=1.0ex
\newdimen\tableaurule\tableaurule=0.4pt
\newdimen\tableaustep
\def\phantomhrule#1{\hbox{\vbox to0pt{\hrule height\tableaurule
width#1\vss}}}
\def\phantomvrule#1{\vbox{\hbox to0pt{\vrule width\tableaurule
height#1\hss}}}
\def\sqr{\vbox{%
  \phantomhrule\tableaustep

\hbox{\phantomvrule\tableaustep\kern\tableaustep\phantomvrule\tableaustep}%
  \hbox{\vbox{\phantomhrule\tableauside}\kern-\tableaurule}}}
\def\squares#1{\hbox{\count0=#1\noindent\loop\sqr
  \advance\count0 by-1 \ifnum\count0>0\repeat}}
\def\tableau#1{\vcenter{\offinterlineskip
  \tableaustep=\tableauside\advance\tableaustep by-\tableaurule
  \kern\normallineskip\hbox
    {\kern\normallineskip\vbox
      {\gettableau#1 0 }%
     \kern\normallineskip\kern\tableaurule}%
  \kern\normallineskip\kern\tableaurule}}
\def\gettableau#1 {\ifnum#1=0\let\next=\null\else
  \squares{#1}\let\next=\gettableau\fi\next}

\newcommand{\be}{\begin{equation}}
\newcommand{\ee}{\end{equation}}
\newcommand{\bea}{\begin{eqnarray}}
\newcommand{\eea}{\end{eqnarray}}
\newcommand{\ba}{\begin{array}}
\newcommand{\ea}{\end{array}}

\newcommand{\id}{\hbox{1\kern-.27em l}}

\newcommand{\RR}{\mathbb{R}}

\newcommand{\ha}{\hat{a}}

\newcommand{\al}{\alpha}
\newcommand{\ga}{\gamma}
\newcommand{\Ga}{\Gamma}
\newcommand{\bet}{\beta}
\newcommand{\vrho}{\varrho}
\newcommand{\ka}{\kappa}
\newcommand{\vka}{\varkappa}
\newcommand{\de}{\delta}
\newcommand{\vphi}{\varphi}
\newcommand{\ep}{\epsilon}
\newcommand{\si}{\sigma}
\newcommand{\la}{\lambda}

\newcommand{\om}{\omega}

\newcommand{\ze}{\zeta}
\newcommand{\De}{\Delta}
\newcommand{\La}{\Lambda}

\newcommand{\Ups}{\Upsilon}

\newcommand{\cN}{\mathcal{N}}
\newcommand{\cO}{\mathcal{O}}
\newcommand{\cW}{\mathcal{W}}
\newcommand{\cF}{\mathcal{F}}
\newcommand{\cT}{\mathcal{T}}
\newcommand{\cV}{\mathcal{V}}

\newcommand{\D}{{\rm d}}
\newcommand{\pa}{\partial}
\newcommand{\rar}{\rightarrow}

\newcommand{\non}{\nonumber}
\newcommand{\lb}{\langle}
\newcommand{\rb}{\rangle}

\newcommand{\SU}{\mathrm{SU}}
\newcommand{\SO}{\mathrm{SO}}

\newcommand{\Sp}{\mathrm{Sp}}

\newcommand{\U}{\mathrm{U}}

\newcommand{\ts}{\textstyle}

\begin{document}

\begin{center}

\vspace*{5mm}
{\Large\sf  $A_{N-1}$ conformal Toda field theory correlation functions \\ from 
conformal $\cN=2$ {\large $\SU(N)$} quiver gauge theories
}

\vspace*{5mm}
{\large Niclas Wyllard}

\vspace*{5mm}
Department of Fundamental Physics\\
Chalmers University of Technology\\
S-412 96 G\"oteborg, Sweden\\[3mm]
{\tt wyllard@chalmers.se}          

\vspace*{5mm}{\bf Abstract:} 
\end{center}
\noindent We propose a relation between correlation functions in the 2$d$ $A_{N-1}$ conformal Toda theories and the Nekrasov instanton partition functions in certain conformal \mbox{$\cN=2$ $\SU(N)$} 4$d$ quiver gauge theories. Our proposal generalises the recently uncovered relation between the Liouville theory and  $\SU(2)$ quivers \cite{Alday:2009}. New features appear in the analysis that have no counterparts in the Liouville case.

\setcounter{equation}{0}
\section{Introduction}

Last month Alday, Gaiotto and Tachikawa uncovered a remarkable connection between a class of conformal $\cN=2$ supersymmetric $\SU(2)$ quiver gauge theories and the conformal Liouville field theory in two dimensions \cite{Alday:2009}. 

The class of $\SU(2)$ quiver theories was introduced and studied in \cite{Gaiotto:2009} (see also \cite{Gaiotto:2009b,Tachikawa:2009,Benini:2009}) and comprise conformal $\cN=2$ theories whose gauge group is a product of $\SU(2)$ factors. The cases with matter in only fundamental, adjoint and bifundamental representations are quivers of conventional type, but the class of theories  also contains more exotic possibilities (called generalized quivers in \cite{Gaiotto:2009}). As emphasised in \cite{Gaiotto:2009} it is convenient to focus on a maximal rank subgroup of the flavour symmetry group, composed only of $\SU(2)$ factors. The quiver can then be drawn in a way so that the $\SU(2)$ flavour factors correspond to external legs.  When drawn in this manner the quivers resemble the diagrams associated with conformal blocks in 2$d$ conformal field theory. In fact this is not a coincidence.  
In \cite{Alday:2009} it was argued that the Nekrasov instanton partition function associated with the gauge theory described by a certain quiver diagram is identical to the conformal block of the corresponding diagram in the Liouville field theory. Furthermore, the integral of the absolute value squared of the full partition function (including also the perturbative pieces) gives rise to a correlation function of primary fields in the Liouville theory. 
Note that in this proposal different correlation functions in {\it one} 2$d$ CFT correspond to instanton partition functions  in  {\it different} 4$d$ gauge theories.  

A natural question to ask is if there are similar connections between the class of 4$d$ $\SU(N)$ $\cN=2$ quiver theories discussed in \cite{Gaiotto:2009} and some class of 2$d$ CFTs. 

The two-dimensional $A_{N-1}$ conformal Toda field theories are a generalisation of the Liouville model (which is identical to the $A_1$ Toda theory). In this paper we  argue that there is a relation between the $\SU(N)$ quiver theories and the $A_{N-1}$ Toda theories. 

In the next section we briefly review the proposal in \cite{Alday:2009} focusing on those aspects that are most relevant for the extension to the class of $\SU(N)$ quiver theories. In section \ref{toda}, we first review the 2$d$ conformal Toda theories as well as the 4$d$ quiver theories with $\SU(N)$ gauge groups and  then make a proposal for how they are connected.  We also perform a few tests of the suggested relations. The proposal shares many features with the Liouville case, but there are also new features on both the CFT and the gauge theory side.
 We conclude with a brief summary and discuss some open problems.  Some more technical aspects have been relegated to an appendix.

\setcounter{equation}{0}
\section{Liouville \& $\SU(2)$ quivers}\label{liou}
In this section we  review the proposal made in \cite{Alday:2009}. This is done in a way which makes it easy to highlight the differences between the Liouville theory and the Toda theories discussed in section \ref{toda}.

\subsection{The Liouville conformal field theory}
The Liouville field theory is defined by the action
\be \label{lact}
S = \int \D^2\si \sqrt{g}\bigg[ \frac{1}{4\pi} g^{ad}\pa_{a}\phi \pa_d \phi + \mu \, e^{2b\phi}+\frac{Q}{4\pi}R \,\phi \bigg] \,,
\ee
where $g_{ad}$ ($a,d=1,2$) is the metric on the two-dimensional worldsheet, and $R$ is its associated curvature. This theory is conformal provided $Q$ and $b$ are related via:
\be
Q=b+\frac{1}{b} \,,
\ee
and the central charge of the theory is 
\be
c = 1 + 6 (b+\frac{1}{b})^2\,.
\ee
The Liouville theory has a set of primary fields 
\be
V_{\al} = e^{2 \al \phi }\,,
\ee
with conformal dimension $\De(\al) = \al(Q-\al)$. 

As is well known, the general form of the three-point correlation function in a 2$d$ conformal field theory is 
\be \label{3pt}
    \lb V_{\al_1}(z_1,\bar{z}_1)
    V_{\al_2}(z_2,\bar{z}_2)V_{\al_3}(z_3,\bar{z}_3)\rb=\frac
    {C(\al_1,\al_2,\al_3)}{ |z_{12}|^{2(\De_1+\De_2-\De_3)} 
    |z_{13}|^{2(\De_1+\De_3-\De_2)} |z_{23}|^{2(\De_2+\De_3-\De_1)}} \,.
\ee
The correlation function of three primary fields in the Liouville theory was calculated in \cite{Dorn:1994} (see also \cite{Teschner:2003}) and takes the form
\bea
    &&C(\al_1,\al_2,\al_3)= 
    \left[\pi\mu\ga(b^2)b^{2-2b^2}\right]^{\frac{Q-\al_1-\al_2-\al_3}{b}} \\ &&\times\,
    \frac{ \Ups(b) \Ups(2\al_1)
     \Ups(2\al_2)  \Ups(2\al_3) }{\Ups(\al_1+\al_2+\al_3 -Q) \Ups(-\al_1+\al_2+\al_3)
\Ups(\al_1-\al_2+\al_3) \Ups(\al_1+\al_2-\al_3)}\,. \non
\eea
The function $\Ups(x)$ is defined as follows (note that $\Ups(x)$ depends on $b$ even though this is not indicated explicitly)
\be \label{ups}
\Ups(x) = \frac{1}{\Ga_2(x|b,b^{-1}) \Ga_2(Q-x|b,b^{-1})   } \,,
\ee
where $\Ga_2(x| \ep_1,\ep_2) $ is defined via the relations
\be \label{Gamma1}
\log \Ga_2(x| \ep_1,\ep_2 ) = \frac{\D}{\D s} \ze_2(s,x|\ep_1,\ep_2 )\Big|_{s=0} \,,
\ee
and 
\be \label{Gamma2}
\ze_2(s,x| \ep_1,\ep_2) = \sum_{m,n}\frac{1}{(m \ep_1 + n \ep_2 +x)^s}  = \frac{1}{ \Ga(s) } \int \D t \,t^{s-1} \frac{e^{-tx} }{ (1-e^{-\ep_1t})(1-e^{-\ep_2t}) } \,.
\ee
The function $\Ga_2(x| \ep_1,\ep_2 )$ satisfies the identity
\be \label{gaid}
 \Ga_2(x+\ep_1| \ep_1,\ep_2 )\,\Ga_2(x+\ep_2| \ep_1,\ep_2 )=x\,\Ga_2(x| \ep_1,\ep_2 )\,\Ga_2(x+\ep_1+\ep_2| \ep_1,\ep_2 )\,.
\ee
Finally, 
\be \label{ga}
\ga(x) = \frac{\Ga(x)}{\Ga(1-x)}\,.
\ee
where $\Ga(x)$ is the ordinary gamma function. 
 The functions $\Ups(x)$ and $\ga(x)$ are related via (note that $\Ups(Q-x)=\Ups(x)$ and $\Ups(b)=\Ups(\frac{1}{b})=\Ups'(0)$)
\bea
\Ups(x+b) &=& \ga(bx)\,b^{1-bx}\,\Ups(x) \,, \non \\
\Ups(x+\frac{1}{b}) &=& \ga(\frac{x}{b})\,b^{2x/b-1}\,\Ups(x) \,.
\eea

Higher-point correlation functions in any CFT can be related to the three-point function of primary fields, which therefore determines the entire theory \cite{Belavin:1984}.  (In subsequent formul\ae{} we suppress the dependence on the antiholomorphic coordinates.) The general form of the four-point function is 
\be \label{4ptgen}
\lb V_{\al_1}(z_1)  V_{\al_2}(z_2) V_{\al_3}(z_3) V_{\al_4}(z_4) \rb  =\left( \prod_{i<j} z_{ij}^{\de_{ij}} \right) F_{\al_1,\al_2,\al_3,\al_4}(z) \,.
\ee
Here $\de_{ij}=\de_{ji}$ is any solution of $\sum_{i\neq j} \de_{ij} = 2\De_j$, $z_{ij}=z_i-z_j$ and $z=z_{12}z_{34}/z_{23}z_{41}$ is the cross ratio. Different choices of $\de_{ij}$ can be absorbed in a redefinition of $ F_{\al_1,\al_2,\al_3,\al_4}(z)$. It is convenient to fix three points to $0, 1, \infty$ and define 
\bea \label{4pt}
\lb V_{\al_1}(0)  V_{\al_2}(1) V_{\al_3}(z) V_{\al_4}(\infty) \rb  &= &\lim_{z_4 \rar  \infty} z_4^{2\De_4}  \lb V_{\al_1}(0)  V_{\al_2}(1) V_{\al_3}(z) V_{\al_4}(z_4) \rb \non
\\ &=& z^{\de_{12} } (1-z)^{ \de_{13} }  F_{\al_1,\al_2,\al_3,\al_4}(z)\,.
\eea
In the Liouville theory we use a bra-ket notation, which has the property $\lb \al|\al \rb=1$, and is such that
\be
\lb \al_1| V_{\al_2}(1) V_{\al_3}(z) |\al_4 \rb = \lb V_{Q-\al_1}(0)  V_{\al_2}(1) V_{\al_3}(z) V_{\al_4}(\infty) \rb\,.
\ee
Inserting a complete set of states we find
\be \label{cb}
\lb \al_1| V_{\al_2}(1) V_{\al_3}(z) |\al_4 \rb = \int \D \al \sum_{{\bf k},{\bf k}'} 
\lb \al_1| V_{\al_2}(1) |\psi_{\bf k}(\al) \rb (K^{-1})_{{\bf k},{\bf k}' } \lb \psi_{{\bf k}'}(\al)| V_{\al_3}(z) |\al_4 \rb\,.
\ee
Here the intermediate states $|\psi_{\bf k}(\al)\rb$ are descendants of the primary state labelled by $\al$, i.e.
\be
|\psi_{-\bf k}(\al) \rb \equiv  L_{-\bf k}  | \al \rb = L_{-k_1} \cdots L_{-k_p} | \al \rb\,,
\ee
and ${\bf k}=(k_1,\ldots,k_p)$ is a partition of $k\equiv|{\bf k}|$ i.e.~$\sum_{i=1}^{p} k_i=k$.  Finally, $K$ is the Gram matrix defined as 
\be
K=\lb \psi_{-\bf k} |\psi_{-\bf k'}\rb\,.
\ee
A special situation arises if one of the $\al_i$ corresponds to a degenerate state  i.e.~a state that is annihilated by some combination of $L_{\bf k}$'s. In such a situation there will be restrictions on the allowed intermediate states, and it may be that only a discrete number is allowed.

The above expression can be represented graphically as 

\begin{center}
\vspace*{0.5 cm}
 \includegraphics{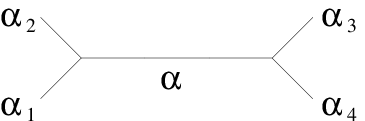}\\[5pt]
{\bf\sf Figure 1:} Conformal block of four-point function in the $s$-channel.\\[20pt]
\end{center}

Note that (\ref{cb}) can be calculated perturbatively using the commutation relations between the $L_{k}$'s and $V_{\al}(z)$\footnote{
Here  $(L_{-1}V_\al)$ is the field obtained by acting with $L_{-1}$ on $V_\al$. Using the known properties of the Virasoro generators gives $(L_{-1}V_\al)=\pa V_\al$.}:
\bea
[L_m, V_{\al}] &=&  z^m[ \,(m+1) \, \De(\al) \, V_\al + z\, (L_{-1} V_\al) \,] \non\\
&=&  z^m( \,m\, \De(\al)\,  V_\al +  [L_0,V_\al]  \,) \,.
\eea  
It is easy to see that $\lb \psi_{{\bf k}'}(\al)| V_{\al_3}(z) |\al_4 \rb$ is proportional to $\lb \al| V_{\al_3}(z) | \al_4 \rb$. Therefore (\ref{cb}) can in principle be determined \cite{Belavin:1984}. The ratio
\be
\frac{\sum_{{\bf k},{\bf k}'}  \lb \al_1| V_{\al_2}(1) |\psi_{\bf k}(\al) \rb (K^{-1})_{{\bf k},{\bf k}' } \lb \psi_{{\bf k}'}(\al)| V_{\al_3}(z) |\al_4 \rb }{ 
\lb \al_1| V_{\al_2}(1) |\al \rb \lb \al| V_{\al_3}(z) |\al_4 \rb }
\ee
is called a conformal block. General $n$-point functions can be dealt with in an analogous manner. They depend on $n-3$ cross ratios.

\subsection{The conformal $\cN=2$ $\SU(2)$ quiver theories } \label{su2}

The class of $\cN=2$ $\SU(2)$ quiver gauge theories introduced and studied in \cite{Gaiotto:2009} have matter (flavour) fields in various representations. Let us recall the following general results: if we have $n$ equal, pseudoreal representations then the global flavour symmetry group contains an $\SO(2n)$ factor; if we have $n$ equal, real representations then we get an $\Sp(2n)$ flavour factor; and if the equal representations are complex we find $\U(n)$ flavour symmetry. 

To get a conformal theory we need to have a suitable matter content to get a vanishing $\bet$-function. If the gauge group is a single $\SU(2)$, we can get a conformal theory from the following matter content: either 4 fundamentals or one adjoint. As the fundamental representation of $\SU(2)$ is pseudoreal the first theory has an $\SO(8)$ flavour symmetry which has an $\SO(4)^2=\SU(2)^4$ subgroup. The adjoint is a real representation so the flavour symmetry of the second theory is $\Sp(2)=\SU(2)$. The resulting two theories can be illustrated graphically in terms of quiver diagrams as in the following figure:
\begin{center}
\vspace*{0.0 cm}
 \includegraphics{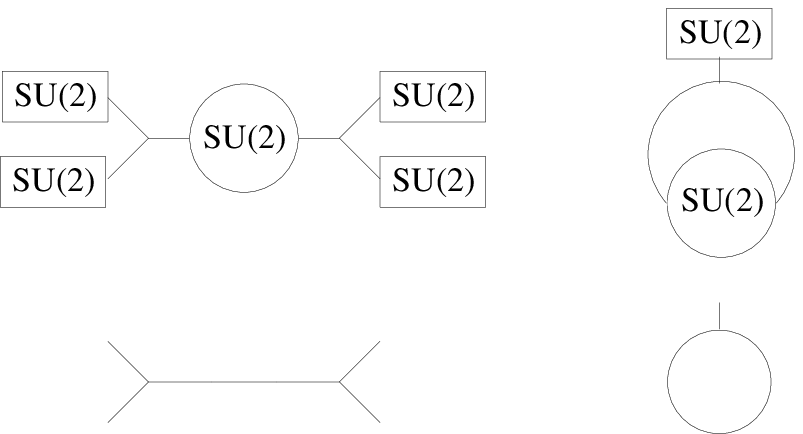}\\[5pt]
{\bf\sf Figure 2:} Quiver diagrams for the two simplest $\SU(2)$ quivers.\\[20pt]
\end{center}

Here boxed $\SU(2)$'s correspond to $\SU(2)$ factors in the flavour symmetry group, whereas circled $\SU(2)$'s correspond to gauge groups.  Simplifying the quiver diagrams by stripping off the boxes and circles we find the diagrams on the second line of the above figure.

We can also consider gauge theories with a product gauge group. In the figure below we draw the quiver diagram for the $\SU(2){\times}\SU(2)$ theory with two matter fields in the fundamental representation of each factor of the gauge group, and one matter field in the  bifundamental representation (which is a real representation):

\begin{center}
\vspace*{0.2 cm}
 \includegraphics{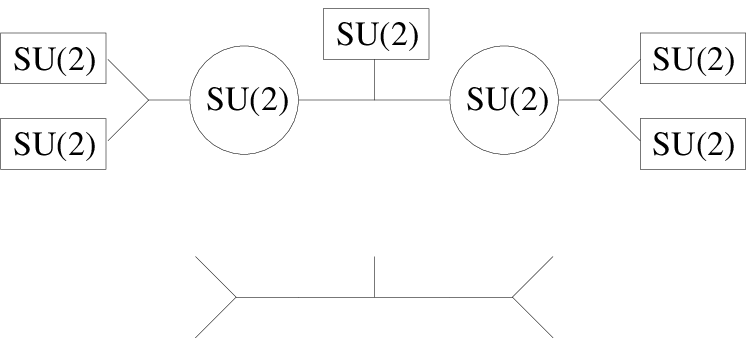}\\[5pt]
{\bf\sf Figure 3:} Another example of an $\SU(2)$ quiver.\\[20pt]

\end{center}

The theories we considered above were all conventional quivers, albeit drawn in a slightly unconventional manner. An important insight in \cite{Gaiotto:2009} was that these theories  belong to a larger class of theories, denoted $\cT_{(n,g)}(A_1)$. The theories in this larger class  can be viewed as the arising from the six-dimensional $A_1$ (2,0) theory \cite{Witten:1995} compactified on $C\times \RR^4$ where $C$ is a genus $g$ Riemann surface with $n$ punctures. The punctures arise from $n$  codimension 2 defects of a certain type which fill $\RR^4$ and intersect $C$ at points. From the gauge theory perspective, the punctures correspond to the $\SU(2)$ factors in the flavour symmetry group, and the genus of the Riemann surface depends on the number of loops in the quiver diagram. It is the $A_1$ theory that is relevant since the gauge group contains $\SU(2)$ factors.  

Consider the $\SU(2)+4\tableau{1}$ theory discussed above. The Riemann surface in this case is a sphere with four punctures.  When this Riemann surface is deformed into two spheres with two punctures each  connected by a thin tube as in the figure below we obtain the weakly coupled gauge theory description (note the obvious similarity with the quiver diagram)

\begin{center}
\vspace*{0.5 cm}
 \includegraphics{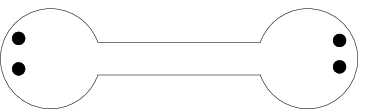} \\[5pt]
{\bf\sf Figure 4:} A particular degeneration of the sphere with four punctures. \\[20pt]
\end{center}

In the second example above ($\SU(2)$ with adjoint matter) the Riemann surface is a torus with one puncture.

There are many more things that could be discussed. Here we only very briefly mention a few further salient points and refer to the original paper \cite{Gaiotto:2009} for more details.   

$\bullet$ The rules for drawing quiver diagrams corresponding to theories in $\cT_{(n,g)}(A_1)$ also involve  an internal three-point vertex. Quivers involving such vertices were called generalised quivers in \cite{Gaiotto:2009}.

$\bullet$ S-duality was an important guideline in the  construction of the $\cT_{(n,g)}(A_1)$ theories. An S-duality transformation in general changes the quiver diagram in a non-trivial way.

$\bullet$ The Riemann surface $C$ is not quite the Seiberg-Witten curve. The Seiberg-Witten curve was constructed in \cite{Gaiotto:2009} (drawing on earlier work \cite{Witten:1997}) and is a certain branched covering of $C$. The Seiberg-Witten differential was also written down. Both the Seiberg-Witten curve and differential were written in a way which clarified the six-dimensional origin of the class of theories. 

$\bullet$ The gauge coupling constants are associated with sewing parameters, $q_i$, of the tubular regions via $q_i=e^{2\pi i \tau_i}$ where $\tau_i$ is the gauge coupling of $i$th factor of the gauge group (more precisely the $\tau$ that appears here is $\tau_{\rm UV}$, cf.~\cite{Grimm:2007,Alday:2009}).

\subsection{The relation between the two}

We saw above that the $\SU(2)$ quiver diagrams when drawn in a particular way (cf.~figures 2, 3) look like the diagrams associated with the conformal blocks in a 2$d$ conformal field theory (cf.~figure 1). 
In \cite{Alday:2009} this resemblance was turned into a concrete proposal relating the $\SU(2)$ quiver theories to a particular conformal field theory --- the Liouville theory. 

In an $\cN=2$ gauge theory the natural basic object to consider is the prepotential. This is most efficiently determined using the instanton counting method of Nekrasov~\cite{Nekrasov:2002,Nekrasov:2003}. In this approach one introduces a deformation of the theory with two parameters $\ep_1$ and $\ep_2$ which belong to an  $\SO(2){\times}\SO(2)$ subgroup of the $\SO(4)$ Lorentz symmetry. The power of this deformation is that it ensures that the integrals over instanton moduli space localise to points; the integrals can therefore be performed in an algorithmic manner. 
The fundamental object in Nekrasov's approach is the partition function\footnote{For simplicity we focus on a quiver with a single gauge group and hence one $a$ and one $q$. Some further information can be found in appendix \ref{appi}. } $Z(a,\mu_j,\ep_1,\ep_2,q)$ where $a$  parameterise the Coulomb branch, the $\mu_j$ are the masses of the matter fields and $q=e^{2\pi i \tau}$ (since we consider a conformal theory the instanton expansion is in terms $q$ and not $\La$). The partition function factorises into two parts as  
\be \label{z}
Z = Z_{\rm pert}\,Z_{\rm inst}\,,
\ee 
where $Z_{\rm pert}$ is the contribution from perturbative calculations (because of supersymmetry there are contributions only at tree- and 1-loop-level), and $Z_{\rm inst}$ is the contribution from the instantons. The instanton part can be expanded as ($k$ is the instanton number) 
\be \label{zinst}
Z_{\rm inst} = \sum_{k} q^{k} Z_{k}  \,.
\ee
 We should stress that one needs the theory to be weakly coupled to be able to apply the instanton counting method.  The $\cN=2$ prepotential $\cF$ is recovered from $Z$ when $\ep_1 = -\ep_2=\hbar$ via the following formula (in the limit $\hbar\rar 0$)
\be
Z = e^{-\frac{1}{\hbar^2}\cF}\,,
\ee
although in this paper it is $Z$ rather than $\cF$ that will be important. 

The proposal in \cite{Alday:2009} is as follows. The instanton partition function (\ref{zinst}) of a certain quiver diagram is conjectured to be equal to the conformal block in the Liouville theory corresponding to the same diagram. The relations between the parameters on the two sides are
\be \label{eb}
\ep_1 = b \,, \qquad \ep_2 = \frac{1}{b}\,,
\ee
the $q_i$ are identified with the cross-ratios $z_i$ and the masses $m_i$ associated with the various $\SU(2)$ factors of the flavour symmetry essentially correspond to the external $\al_i$ in the Liouville theory. The $a$'s of the $\SU(2)$ gauge factors essentially correspond to the internal $\al$'s in the conformal block. (The precise rules will be illustrated  in a simple example below).

We saw above that a quiver diagram with $n$ external legs corresponds to a particular degeneration of a Riemann surface with $n$ punctures.  In the conformal block this corresponds to choosing a specific intermediate channel (e.g.~the $s$-channel as in figure 1). Other channels in the conformal block correspond to other degenerations of the Riemann surface. The various possibilities are related by the crossing symmetry on the CFT side and by S-duality on the gauge theory side \cite{Alday:2009}. Note that the identification (\ref{eb}) implies that one can not set both $\ep_i$ to zero (since $\ep_1\ep_2$=1); the deformation on the gauge theory side is therefore essential.

A further proposal was also made in  \cite{Alday:2009}\footnote{This proposal was inspired by the earlier work \cite{Pestun:2007}.}: the absolute value squared of the complete partition function (\ref{z}) integrated over the $a$'s of the $\SU(2)$ gauge factors with the natural measure should correspond to the full Liouville correlation function. Essentially what happens is that the perturbative contributions combine to give the three-point function factors and the instantons give the conformal block. 

A simple example will illustrate the proposal. Consider the $\SU(2)+4\tableau{1}$ theory (the first example in figure 2). For this theory the $Z_k$  can be calculated in two ways. One can either view $\SU(2)$ as a restriction of $\U(2)$ or as $\Sp(2)$. In the latter case one can use the results for $\Sp(2n)$ instanton counting obtained in \cite{Marino:2004,Nekrasov:2004}. The leading results obtained using the two methods are listed in (\ref{suninun}), (\ref{su2insp2}). We find that the two expressions are related as 
\be \label{u(1)}
Z^{\SU(2)\subset \U(2)}_{\rm inst}(a,m+\frac{\ep}{2},\ka-\ep) = (1-q)^{Q_2(a,m,\ka,\ep) } Z^{\SU(2)=\Sp(2)}_{\rm inst}(a,m,\ka)\,,
\ee
where $Q_2(a,m,\ka,\ep)$ is the quadratic expression
\be
Q_2(a,m,\ka,\ep) =  \frac{1}{2}(a^2-m_1^2 -m_2^2  +\ka_1^2 +\ka_2^2+4\ka_1\ka_2-2\ka_1 \ep -2\ka_2 \ep) - \frac{3}{8}\ep^2  \,.
\ee
In  \cite{Alday:2009} it was argued that one should factor out an overall $(1-q)^{Q'_2(\ka,\ep)}$ from $Z^{\SU(2)\subset \U(2)}_{\rm inst}$ when comparing with the CFT conformal block. Here $Q_2'$ is a certain quadratic expression, different from $Q_2$, that only depends on $\ka_i$ and $\ep$.  It was argued that this prefactor was due to an incomplete decoupling of the $\U(1)$ inside $\U(2)$. 
The prepotential of an $\cN=2$ gauge theory is only determined up to a quadratic expression, which at the level of the partition function translates into the statement that $\log Z$ is only defined up to a quadratic expression. We see that the different prefactors are examples of this ambiguity and therefore do not lead to a contradiction. Since the prefactors always seem to involve $(1-q)$ it is not inconceivable that the different calculations correspond to different choices of the $\de_{ij}$ in (\ref{4pt}).

The prefactor one has to factor out from $Z^{\SU(2)\subset \U(2)}_{\rm inst}$ to get agreement with the conformal block calculation is closely related to $\U(1)$ in the following sense. If we directly calculate the partition function for a $\U(1)$ theory coupled to a fundamental and an anti-fundamental matter field we find 
\bea
Z_{\rm inst} = \sum_{Y_{\bf k}} q^{|{\bf k}|} \prod_{{\rm boxes}\in Y_{\bf k}} \frac{ (\ep_1\,i + \ep_2 \,j -\mu )(\ep_1 \,i + \ep_2 \,j  + \bar{\mu} - \ep)}{( \ep_2 (A {+}1) -\ep_1 L )(\ep - \ep_2 (A {+}1) + \ep_1 L)    } = (1-q)^{\frac{(\mu-\ep)\,\bar{\mu}}{\ep_1\ep_2}} 
\eea
where the sum runs over all Young diagrams and the product is over all boxes in the diagram; $(i,j)$ label the columns and rows and $A$ and $L$ are the arm and leg lengths of a box. 
A similar calculation for the second model in figure 2 gives (cf.~(6.12) in \cite{Nekrasov:2003})
\be 
 \sum_{k} q^k \prod_{{\rm boxes}} \frac{ ( \ep_2 (A {+}1) -\ep_1 L-\mu )(\ep - \ep_2 (A {+}1) + \ep_1 L -\mu) }{( \ep_2 (A {+}1) -\ep_1 L )(\ep - \ep_2 (A {+}1) + \ep_1 L)    }     = \prod_{n=1}^{\infty}(1-q^n)^{\frac{\mu\,(\ep-\mu)}{\ep_1\ep_2}-1} \,.
\ee
The above two expressions (with $\ep_1\ep_2=1$) are similar to (C.1), (C.5) in \cite{Alday:2009}.

Continuing with our example, we calculate the first non-trivial term in the conformal block as
\bea
&&\frac{\lb \al_1| V_{\al_2}(1) L_{-1}|\al \rb ( \lb \al| L_{1}  L_{-1}|\al \rb )^{-1} \lb \al| L_1 V_{\al_3}(z) |\al_4 \rb}{\lb \al_1| V_{\al_2}(1) |\al \rb  \lb \al| V_{\al_3}(z) |\al_4 \rb}  \non \\ &=& z\,\frac{[\De(\al) +\De(\al_2) - \De(\al_1)][\De(\al) +\De(\al_3)- \De(\al_4)] }{2\De(\al) }\,.
\eea
Comparing this expression to the one-instanton calculation in the $\SU(2)$ theory (\ref{suninun}) we find agreement provided we factor out $(1-q)^{(2\ka_1 + Q))(2\ka_2 + Q)/2}$ from $Z$ and identify (in our conventions)
\be \label{match}
z=-q\,,\;\; m_1= \al_1\,,\;\;m_2= \al_4\,,\;\; \ka_1 = -\al_2-Q/2 \,,\;\; \ka_2 =-\al_3-Q/2   \,,\;\; a = \al-Q/2  \,.
\ee

Furthermore, using (\ref{zpert}) we see that the perturbative contribution to the partition function is 
\bea
&&Z_{\rm pert} = \exp\left[ -\ga_{\ep_1\ep_2} (2a-\ep_1) -\ga_{\ep_1\ep_2} (2a-\ep_2) +  \ga_{\ep_1\ep_2} (-a-m_1-\ka_1)\right]  \\[3pt]  &&\!\!\!\!\!\!\!\!\! \cdot\, \exp\left[\ga_{\ep_1\ep_2} (-a-m_2- \ka_2) + \ga_{\ep_1\ep_2} (-a+m_1-\ka_1-\ep) + \ga_{\ep_1\ep_2} (-a+m_2-\ka_2-\ep )\right] \non \\[3pt] &&\!\!\!\!\!\!\!\!\! \cdot\, \exp\left[\ga_{\ep_1\ep_2} (a{-}m_1{-}\ka_1){+}\ga_{\ep_1\ep_2} (a{-}m_2{-} \ka_2) {+} \ga_{\ep_1\ep_2} (a{+}m_1{-}\ka_1{-}\ep) {+} \ga_{\ep_1\ep_2} (a{+}m_2{-}\ka_2{-}\ep )\right] \non
\eea
whose absolute value squared should be compared to (via (\ref{gag}), (\ref{ups}) and the identifications (\ref{eb}) and (\ref{match}))
\bea
&& \lb \al_1| V_{\al_2}(1) |\al \rb  \lb \al| V_{\al_3}(z) |\al_4 \rb = \left[ \pi\mu\ga(b^2)b^{2-2b^2}\right]^{(\al_1 -\al_2 -\al_3-\al_4)/b}  \\
&&\!\!\!\!\!{\times}\, \frac{\Ups(b)\Ups(2(Q-\al_1))\Ups(2\al_2)\Ups(2\al)}{\Ups(-\al_1+\al_2+\al)\Ups(Q-\al_1+\al_2-\al)\Ups(Q-\al_1-\al_2+\al)\Ups(-Q+\al_1+\al_2+\al)}\non \\ 
&&\!\!\!\!\!{\times} \, \frac{\Ups(b)\Ups(2(Q-\al))\Ups(2\al_3)\Ups(2\al_4)}{\Ups(-\al+\al_3+\al_4)\Ups(Q-\al+\al_3-\al_4)\Ups(Q-\al-\al_3+\al_4)\Ups(-Q+\al+\al_3+\al_4)}.\non 
\eea

If we redefine the primary fields $V_\al$ by introducing 
\be 
\cV_{\al} =\frac{ \left[ \pi\mu\ga(b^2)b^{2-2b^2}\right]^{\al/b} }{\Ups(2\al)} V_{\al}\,,
\ee
and use the identity (\ref{gaid}) most of the differences between $\int \D a\, (2a)^2 \, |Z|^2$ and the four-point correlation function can be removed. The remaining discrepancy is a factor depending only on $b$, as well as the $q$ and $(1-q)$ pieces. It does not seem unlikely that these remaining differences can also be removed by a proper definition of $Z$. In this context we note that the $(1-q)$ factors  one naturally gets in the instanton calculations  are not unsimilar to those that have appeared in the Liouville literature, see e.g.~(2.7) in \cite{Fateev:2009}. Factors of $q^{Q_2''}$ can be obtained from (unphysical) $2\pi i \tau Q''_2$ pieces in the prepotential.

\setcounter{equation}{0}
\section{Toda \& $\SU(N)$ quivers} \label{toda}
In this section we first review the conformal $A_{N-1}$ Toda theories and then review the class of $\SU(N)$ quiver theories. Finally, we make a proposal for how the two are related.

\subsection{The $A_{N-1}$ conformal Toda field theories}
The $A_{N-1}$ Toda field theories are defined by the action
\be \label{tact}
S = \int \D^2\si \sqrt{g}\left[  \frac{1}{8\pi} g^{ad}\lb \pa_{a}\phi, \pa_d \phi \rb + \mu \sum_{i=1}^{N-1} e^{b\lb e_i,\phi\rb} + \frac{\lb Q, \phi \rb}{4\pi}R \,\phi  \right],
\ee 
where $g_{ad}$ ($a,d=1,2$) is the metric on the two-dimensional worldsheet, and $R$ is its associated curvature. Furthermore, the $e_i$ are the simple roots of the $A_{N-1}$ Lie algebra, $\lb \cdot,\cdot\rb$ denotes the scalar product on the root space,  and the $(N{-}1)$-dimensional vector of fields $\phi$ can be expanded as $\phi = \sum_i \phi_i e_i$.

From the above action it is easy to see that the Liouville theory (\ref{lact}) is identical to the $A_1$ Toda field theory\footnote{Because of the standard Lie algebra conventions, some formul\ae{} differ at first sight.}.  The Toda theories with rank ${>}1$ are much more complicated than the Liouville model. Toda theories can be defined for any simple Lie algebra by taking the $e_i$ to be the simple roots of the corresponding Lie algebra\footnote{Toda theories can also be defined for affine Lie algebras by adding an additional simple root. The resulting theories are non-conformal and will play no role in this paper.}. 

The $A_{N-1}$ Toda theory is conformal provided $Q$ and $b$ are related via:
\be \label{Q}
Q = (b+\frac{1}{b})\rho \,,
\ee
where $\rho$ is the Weyl vector (half the sum of all positive roots). The central charge is \cite{Mansfield:1982}\footnote{As written this formula is true only for the $A_{N-1}$ Lie algebras. In the general case (including also non-simply laced cases) the formula gets replaced by  $c = r  + 12 \lb Q, Q\rb $ where $r$ is the rank and $Q = b \rho  + \frac{\rho^\vee}{b}$ where $\rho^{\vee}$ is the dual Weyl vector (half the sum of the positive coroots). Note the duality under $b\leftrightarrow \frac{1}{b}$ and $\rho \leftrightarrow \rho^\vee$.}
\be \label{ct}
c = N - 1 + 12 \lb Q, Q\rb = (N-1)(1+N(N+1) (b+\frac{1}{b})^2)\,,
\ee
where in the second step we used the Freudenthal-de Vries strange formula. 

For $N>2$ the symmetry algebra of the $A_{N-1}$ Toda theories contains in addition to the stress tensor, $T$, also  $N{-}2$ additional holomorphic symmetry currents with conformal dimensions $3,4,\ldots,N{+}1$ \cite{Bilal:1988a}. These symmetry currents are usually  denoted $\cW^{(k)}$ ($k=2,\ldots,N{+}1$), where $\cW^{(2)}\equiv T$, and  form a $\cW_{N+1}$-algebra (see e.g.~\cite{Bouwknegt:1992} for a review of $\cW$-algebras). As an example,  in the $A_2$ Toda theory  there is only a single extra current, $\cW^{(3)}(z)$. Together with the stress tensor it forms the so called $\cW_3$ algebra~\cite{Zamolodchikov:1985}. The mode expansions of the currents are 
\be \label{modes}
T(z) = \sum_n L_n z^{-n-2} \,, \qquad \cW^{(3)}(z) = \sum_n W^{(3)}_n z^{-n-3} \equiv \sum_n W_n z^{-n-3}\,.
\ee
In terms of the modes (\ref{modes}) the $\cW_3$ algebra can be written
\bea
{}[L_n,L_m]&=&(n-m)L_{n+m}+\frac{c}{12}(n^3-n)\de_{n,-m}\,, \non 
\\
{}[L_n,W_m]&=&(2n-m)W_{n+m}\,,
\\
{}[W_n,W_m]&=&\frac{c}{3\cdot5!}(n^2-1)(n^2-4)n
    \de_{n,-m}+\frac{16}{22+5c}(n-m)\La_{n+m} \non 
\\ 
&+& \!\!\! (n-m)\left(\frac{1}{15}(n+m+2)(n+m+3)-\frac{1}{6}(n+2)(m+2)
    \right) L_{n+m}\,, \non 
\eea
where $c$ is the central charge (\ref{ct}) and 
\be
 \La_n=\sum_{k=-\infty}^{\infty}:L_kL_{n-k}:+\frac{1}{5}x_n
 L_n\,,
\ee
with
\be
  x_{2l}=(1+l)(1-l)\,,\qquad x_{2l+1}=(2+l)(1-l)\,.
\ee

Primary fields can be defined in analogy with the Virasoro case. A $\cW$-primary field satisfies
\be
W^{(k)}_0 V = w^{(k)} V  \,, \qquad W^{(k)}_{n} V = 0\quad \mathrm{when} \!\!\!\quad n>0\,. 
\ee
In the Toda theories the ($\cW$) primary fields are 
\be \label{prim}
V_{\alpha}=e^{\lb\alpha,\phi\rb}\,.
\ee
In the particular example of the $A_2$ theory the primary fields satisfy 
\be
   L_0 V_{\al} = \De(\al) V_{\al}\,, \qquad W_0 V_{\al} = w(\al) V_{\al}\,,
   \qquad L_{n} V_{\al} = W_{n} V_{\al} = 0 \quad \mathrm{when}\quad\!\!\! n>0 \,,
\ee
where
\be \label{De}
  \De(\al)=\frac{(2Q-\al,\al)}{2}\,,
\ee
is the conformal dimension and
\be \label{w}
  w(\al)=i\sqrt{\frac{48}{22+5c}}\; (\al-Q,\la_1)(\al-Q,\la_2)(\al-Q,\la_3) \,,
\ee
is the quantum number of the $\cW^{(3)}$ current (here the $\la_i$ are the weights of the fundamental representation of the $A_2$ Lie algebra).

The three-point function of primary fields is defined as in (\ref{3pt}).
Except in the $A_1$ case, only partial results are known.  Recently it was shown~\cite{Fateev:2005,Fateev:2007b} that  in the special case when one of the $\al_i$'s takes one of the two special values 
\be \label{spec}
\al = \vka \om_{1} \,, \quad\mathrm{or} \quad \qquad \al = \vka \om_{N-1}\,,
\ee
where $ \om_{1}$ ($\om_{N-1}$) is the highest weight of the fundamental (antifundamental) representation of the $A_{N-1}$ Lie algebra, the three-point function is given by  (\ref{3pt}) with
\bea \label{t3pt}
    C(\al_1,\al_2,\vka \omega_{N-1})&=&
    \left[\pi\mu\ga(b^2)b^{2-2b^2}\right]^{\frac{\lb 2Q-\sum\al_i,\rho \rb}{b}}\\ &
   \times&\!\!\! \frac{\left(\Ups(b)\right)^{N-1}\Ups(\vka)\prod_{e>0}\Ups\big(\lb Q-\al_1,e\rb\big)
    \Ups\big(\lb Q-\al_2,e\rb\big)}{\prod_{ij}\Ups\big(\frac{\vka}{n}+
    \lb\al_1-Q,\la_i\rb+\lb\al_2-Q,\la_j \rb\big)}\,, \non
\eea
where the product in the numerator is over all positive roots and in the denominator the $\la_i$ are the weights of the representation with highest weight $\om_{N-1}$. (The result for $\al_3=\vka \om_1$ is obtained by replacing $\la_i$ by $\la'_i = -\la_{N+1-i}$.) 

We now come to a crucial difference with the Liouville case. It is no longer true that the higher-point functions of $\cW$ primary fields are determined in terms of the three-point function of $\cW$ primary fields \cite{Bowcock:1993}. 
If one forgets about the full $\cW$-algebra, it is of course still true that the higher-point functions are determined in terms of three-point functions of the {\em Virasoro} primaries (a larger set than the $\cW$ primaries). However,  the $\cW$ symmetry, while constraining, is not powerful enough to determine the higher-point functions of $\cW$ primaries  in terms of the three-point functions of $\cW$ primaries. 
 To illustrate why this is so we focus on the $A_2$ theory. As in (\ref{cb}), the four-point function of ($\cW$) primaries (\ref{prim}) can be decomposed by inserting a complete set of intermediate (descendant) states, which now are given by
\be
|\psi_{({\bf k},{\bf l})}(\al)\rb \equiv L_{-{\bf k}} W_{-{\bf l}}  |\al \rb = L_{-k_1} \cdots L_{-k_p} W_{l_1} \cdots W_{-l_q} |\al \rb\,.
\ee

Using the commutation relations 
\bea
{}[L_m, V_{\al}] &=&  z^m[ \,(m+1) \, \De(\al) \, V_\al + z\, (L_{-1} V_\al) \,]\,, \\
{}[W_m,V_\al] &=&  z^m [\, \frac{ (m+1)(m+2)}{2}\, w(\al) \, V_\al + z\,(m+1)\,  (W_{-1} V_\al) + z^2\, (W_{-2} V_\al) \,] \,, \non
\eea 
one can in general only reduce $\lb \psi_{{\bf k}'}(\al)| V_{\al_3}(z) |\al_4 \rb$ to belong to the set of states 
\be \label{w1n}
\lb \al| V_{\al_3}(z) (W_{-1})^n | \al_4 \rb\,, 
\ee
where $n$ is any positive integer (including zero). It is therefore not true that all correlations functions are determined solely in terms of the three-point functions of $\cW$ primary fields (see \cite{Bowcock:1993} for more details).

However, if some of the $\al_i$ take special values then the corresponding states may be  semi-degenerate i.e.~$ (W_{-1})^p | \al_i \rb$ can be expressed in terms of states of the form  $L_{\bf k} (W_{-1})^n | \al_i \rb$ with $n<p$. In such a situation, one can reduce any $\lb \psi_{{\bf k}'}(\al)| V_{\al_3}(z) |\al_4 \rb$ to  belong to the set (\ref{w1n}) with $n<p$.  In the particular case when $p=1$ the class of higher-point correlation functions are therefore determined in terms of three-point functions of $\cW$ primary fields. 

Continuing with the $\cW_{3}$ example, it is known that \cite{Fateev:1987}
\be \label{deg}
(W_{-1} - \frac{3w(\al)}{2\De(\al)} L_{-1})|\al \rb = 0\,,
\ee
 provided that $\al$ takes one of the two values (\ref{spec}) with $N=3$. The corresponding state is therefore (semi-)degenerate. 
Note that the condition (\ref{deg}) implies (by acting with $W_1$)
\be \label{cond}
\De_\al \left[\frac{32}{22+5c}(\De_\al +\frac{1}{5}) - \frac{1}{5}\right] = \frac{9}{2}\frac{w_\al^2}{\De_\al}\,.
\ee
 
 Denoting the primaries corresponding to states satisfying (\ref{deg}) as $V'_{\vka}$, it follows from the above discussion that $n$-point functions of the form\footnote{Here $\lb \al_1|\cO  |\al_n \rb \equiv \lb V_{2Q-\al_1} \cO\, V_{\al_n}\rb$.}
\be \label{multi}
\lb \al_1| V'_{\vka_2} \cdots  V'_{\vka_{n-1}} |\al_n\rb\,,
\ee
can be calculated perturbatively in terms of $\cW$  chiral blocks (analogous to the Virasoro conformal blocks) and the special class of three-point functions of primary fields given in (\ref{t3pt}).

\subsection{The conformal $\cN=2$ $\SU(N)$ quiver theories }

We now turn to the class of conformal $\cN=2$ 4$d$ $\SU(N)$ quiver theories. As in the $\SU(2)$ case discussed in section \ref{su2}  we start with some simple examples. Theories with a single $\SU(N)$ gauge factor  are conformal if the matter content comprise either  $2N$ fundamentals or one adjoint. In the first case, the flavour symmetry group is $\U(2N)$ since the fundamental representation is complex (for $N>2$). It is convenient to focus on a $\U(N){\times}\U(N)\cong \U(1)^2\SU(N)^2$  subgroup of $\U(2N)$. The adjoint representation is complex so the flavour symmetry of the second model is $\U(1)$. For these models we can draw the following quiver diagrams

\begin{center}
\vspace*{0.2 cm}
 \includegraphics{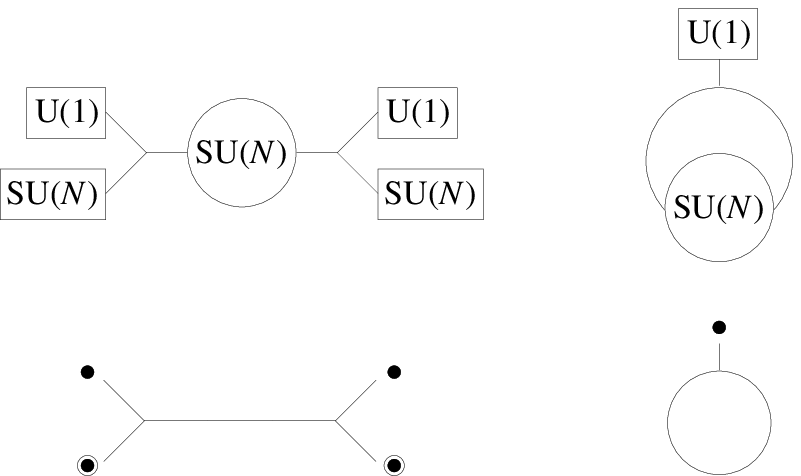}\\[5pt]
{\bf\sf Figure 5:} The two simplest examples of $\SU(N)$ quivers.\\[15pt]
\end{center}

In figure 5, the boxes refer to factors of the flavour symmetry group and the circles to the $\SU(N)$ gauge group.  There are two different types of flavour symmetry factors in the above diagrams in contrast to the the $\SU(2)$ case where  there was only one. In the second line of the above figure we have stripped off the boxes and circles, and used a filled dot and a circled filled dot to indicate the two different types of external legs corresponding to the different flavour symmetries. 

The example in the following figure involves also a bifundamental representation (which is a complex representation for $N>2$) 

\begin{center}
\vspace*{0.3 cm}
 \includegraphics{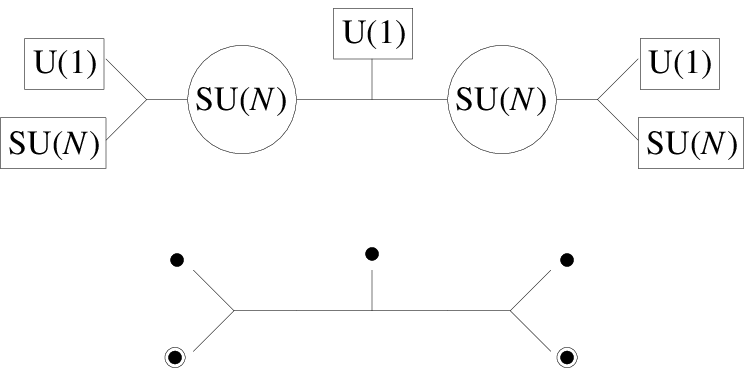}\\[5pt]
{\bf\sf Figure 6:} Another example of an $\SU(N)$ quiver.\\[0pt]
\end{center}

As for the $\SU(2)$ theories, the above examples are particular members of a larger class of theories, denoted $\cT_{(n,g)}(A_{N-1})$ \cite{Gaiotto:2009}.
 This class of theories arise from the six-dimensional $A_{N-1}$ (2,0) theory \cite{Witten:1995} compactified on $C\times \RR^4$ where $C$ is a genus $g$ Riemann surface with $n$ punctures. The punctures are due to  codimension 2 defects filling $\RR^4$ and intersecting $C$ at points. An important difference compared to the $\SU(2)$ case is that now there is more than one type of possible defects. It was argued in \cite{Gaiotto:2009} that the different codimension 2 defects are classified by partitions of $N$ (which can be represented graphically in terms of Young tableaux). The outcome is that one can associate a Young tableau to each puncture. In the above examples we only encountered two kinds of punctures. 

As for the $\SU(2)$ quivers, the genus of the Riemann surface depends on the number of loops in the quiver diagram. The first quiver in figure 5 corresponds to a four-punctured sphere that has degenerated as in figure 4, where each of the two spheres has two punctures, one of each kind. Other distributions of the punctures lead to more exotic descriptions; the different possibilities are related by S-duality, the prototypical example being~\cite{Argyres:2007}. To construct more general quivers, the basic building block is the theory associated with three generic punctures on a sphere. See \cite{Gaiotto:2009} for further details.

\subsection{The relation between the two}

In order to find a relation between the $A_{N-1}$ Toda theories and the $\SU(N)$ quiver theories we need to find a rule that associates to a puncture described by a certain Young diagram a corresponding primary $V_{\al}$ in the Toda theory. In other words, we need a map from the Young tableau to some $\al$. We propose that this map is essentially the same as the one used in \cite{Gaiotto:2009} to associate a set of masses with a puncture. In particular this means (cf.~section 4.2 in \cite{Gaiotto:2009}) that a full puncture is associated with an unconstrained $\al$ whereas the puncture with $\U(1)$ flavour symmetry maps to an $\al$ of the form
\be \label{u1pun}
\bullet \quad \mapsto \quad \qquad \vka \om_1 \qquad \mbox{or} \qquad  \vka \om_{N-1}\,.
\ee
Here $ \om_1$ ($\om_{N-1}$) is the highest weight of the fundamental (antifundamental) representation of $\SU(N)$. To see this we note that in the convention where the $\SU(N)$ root space is spanned by vectors whose components sum to zero, the weights of the fundamental representation of $\SU(N)$  can be chosen to be 
\be \label{weights}
\la_i = -u_i + \frac{1}{N}\sum_{j=1}^{N} u_j  \,,
\ee
where $u_i$ is the vector whose $i$th entry is $1$ with all other components equal to $0$ (note that $\sum_{i=1}^{N}\la_i=0$). The highest weight is
\be \label{om1}
\om_{1} = \la_1 = \frac{1}{N}(1{-}N,1,\ldots,1)\,,
\ee
and the highest weight of the antifundamental representation (with weights $\bar{\la}_i = -\la_{N+1-i}$) is 
\be \label{omn}
\om_{N-1} = -\la_N = \frac{1}{N} (1,\ldots,1,1{-}N)\,.
\ee
Note that in this notation the simple roots are $e_i=u_i-u_{i+1}$ ($i=1,\ldots,N{-}1$) and the positive roots are $u_i-u_j$ ($i<j$). 

Even before doing any calculations we can perform some consistency checks of our proposal. It is not too difficult to see that the cases in which we have a perturbative description in terms of a conventional gauge theory quiver precisely correspond to the correlation functions for which the $\cW$ algebra ambiguities in higher-point correlation functions are absent. This fact is quite encouraging and supports our proposal.

Additional evidence for the rule (\ref{u1pun}) can be obtained by considering the perturbative contributions. The one-loop matter contribution to the prepotential in the $\SU(N)$ theory with $2N$ (anti)fundamentals is proportional to
\be \label{one}
\sum_{\la} \sum_{\varrho\in \mathrm{reps}}(\lb a,\la\rb + \mu_{\varrho})^2 \big[\log(\lb a,\la\rb + \mu_{\varrho})  - {\ts \frac{3}{2}} \big] \,.
\ee
Here the first sum runs over the weights $\la_i$ of the representation $\varrho$ and the second sum runs over the various representations $\vrho$ of the gauge group that the matter fields transform in. In the $\SU(N)$ case the different $\vrho$ comprise $2N$ (anti)fundamental representations.  Decomposing the flavour symmetry as $\SU(N)^2\U(1)^2$ it is natural to take  $N$ masses associated with the matter fields in the  fundamental  representation of $\SU(N)$ (gauge) to transform in the fundamental of (flavour) $\SU(N)$ and similarly for the remaining $N$ matter fields.  We use the conventions  
\be
\mu_i = \frac{\ka}{N} - \lb m, \la_i \rb\,,
\ee 
and 
\be
\bar{\mu}_i=-\frac{\bar{\ka}}{N}  + \lb \bar{m}, \la_i \rb\,,
\ee 
where $\la_i$ are the weights (\ref{weights}). Note that $\sum_i \la_i =0$, so $m$/$\bar{m}$ only contains $N-1$ independent parameters.

The above decompositions means that (\ref{one}) can be written as:
\bea
&&\sum_{i,j=1}^N (\lb a,\la_i \rb - \frac{\ka}{N}  + \lb m,\la_j \rb)^2 [\log(\lb a,\la_i\rb -  \frac{\ka}{N}  + \lb m,\la_j \rb )  - {\ts \frac{3}{2}}] \\
&+& \sum_{i,j=1}^N (\lb a,\la_i \rb - \frac{\bar{\ka}}{N}  + \lb \bar{m},\la_j \rb)^2 [\log(\lb a,\la_i\rb - \frac{\bar{\ka}}{N}  + \lb \bar{m},\la_j \rb )  -{\ts \frac{3}{2} }]\,.
\eea

Note that since both the gauge and flavour groups involve the fundamental representation of $\SU(N)$ both sums run over the same set of weights. This means that the gauge parameters $a$ and the mass parameters $m$, $\bar{m}$ are treated on the same footing. This is required for the identification with a CFT correlation function and provides evidence for the identification of the pieces involving $\ka$ and $\bar{\ka}$ with the $\U(1)$ punctures. 

Including also the $\ep$ dependence using the rules (\ref{zpert}), the perturbative contribution to $Z$ in the $\SU(N)$ theory with $2N$ fundamentals can be written (note that $\prod_{i<j} (a_i-a_j) = \prod_{e>0} \lb a,e \rb$). 
\be
\frac{ \prod_{e>0}\Ups(\lb a,e \rb-\ep_1 )\Ups(\lb a,e \rb -\ep_2)}{ \prod_{i,j} \Ups(\lb a,\la_i \rb  -  \frac{\ka}{N}  + \lb m,\la_j \rb) \Ups(\lb a,\la_i \rb -  \frac{\bar{\ka}}{N}   + \lb \bar{m} ,\la_j \rb -\ep) }\,.
\ee
Comparing this to the contribution in the Toda theory four-point function:
\be
\lb \al_1| V'_{\vka} |\al\rb \lb \al| V'_{\bar{\vka}} |\al_4\rb\,,
\ee 
using (\ref{t3pt}) we find that if we redefine the primaries as
\be \label{newV}
\cV_{\al}  =\frac{ \left[ \pi\mu\ga(b^2)b^{2-2b^2}\right]^{\lb \al,\rho\rb/b} }{\prod_{e>0}\Ups(\lb Q-\al,e\rb)} V_{\al}\,, \qquad \cV'_{\vka}  =\frac{ \left[ \pi\mu\ga(b^2)b^{2-2b^2}\right]^{\lb \vka,\rho\rb/b} }{\Ups(\vka)} V'_{\vka}\,,
\ee
the two expressions agree up to the expected Vandermonde determinant $\prod_{i<j} (a_i-a_j)^2$, and a factor that only depends on $b$, provided the parameters on the two sides are identified as in (\ref{Tmatch}) below. In (\ref{newV}) (and elsewhere) we are using a slight abuse of notation, using $\vka$ to denote both the vector (\ref{u1pun}) and the  coefficient in front of $\om_1$/$\om_{N-1}$ in that vector. It should be clear from the context which is meant. As in the Liouville case it seems possible that the difference between
\be
\int \prod_i \D a_i \prod_{i<j} (a_i-a_j)^2 \, |Z|^2
\ee
and the four-point function can be removed by a minor redefinition of $Z$.

It is fairly straightforward to extend this discussion to any conventional quiver such as  for instance the one in figure 6. Such linear quivers correspond to correlation functions of the type (\ref{multi}). Correlation functions on the torus involving a string of $V'_{\vka}$'s correspond to conventional necklace quivers and can therefore also be studied.

Next we would like to evaluate the instanton corrections on the gauge theory side and compare them to the chiral blocks in the Toda theory. In (\ref{suninun}) the one-instanton correction in the $\SU(N)$ theory with 2$N$ fundamentals is given and in (\ref{block}) we have calculated the chiral block in the $A_2$ Toda theory. Using (\ref{De}) and (\ref{w}), it can be shown that  (\ref{block}) agrees with  (\ref{suninun}) provided we factor out a $(1-q)^{\ka(\bar{\ka}-\ep)/3}$ term from $Z_{\rm inst}$ and identify\footnote{There appears to be more than one possible identification, which is probably due to the symmetry under the Weyl group. Also, there are some differences depending on which choice is made in (\ref{u1pun}) and which weights are used.}
\be \label{Tmatch}
z=-q\,,\;\;   \al_1 = m + Q \,,\;\;  \vka= \ka + 3 Q \,,\;\;  \bar{\vka} = \bar{\ka}+6 Q  \,,\;\;  \al_4 = \bar{m}+Q   \,,\;\;  \al = a + Q  \,.
\ee
This represents a highly non-trivial test of the proposed relation. 
 
We should stress that, if the conjectured relation is correct, the instanton counting gives closed expressions for any correlation function in the class (\ref{multi}) in the $s$ channel. Such expressions would represent a new result. Note also that the intermediate states/descendants on the CFT side are labelled by sets of partitions and there may be a more direct correspondence with the partitions appearing in the instanton counting method.
Finally, we mention that one can also discuss the Seiberg-Witten curve along the lines of \cite{Alday:2009} by replacing the stress tensor with a more general $\cW$ algebra current. 

\setcounter{equation}{0}
\section{Discussion and outlook}

In this paper we argued that the connection proposed in \cite{Alday:2009} between the Liouville theory in two dimensions and 4$d$ $\cN=2$ $\SU(2)$ quiver theories, extends to a connection between the 2$d$ $A_{N-1}$ Toda theories and the class of 4$d$ $\cN=2$ $\SU(N)$ quiver theories studied in \cite{Gaiotto:2009}. Although we have only performed selected tests of this idea, the agreement is nevertheless quite striking. 

It is clearly important to perform further tests of the suggested relations. One possible approach is the following. It has been shown that when some of the $\al_i$ take certain special values, the four-point correlation functions (and also higher-point functions) in the Liouville \cite{Fateev:2007a,Fateev:2009} and Toda theories \cite{Fateev:2005,Fateev:2007b,Fateev:2008} satisfy certain differential equations. These equations can be solved exactly in terms of special functions with explicit integral representations. On the gauge theory side it might be possible to sum up all instanton contributions and make contact with the results in  \cite{Fateev:2007a,Fateev:2009,Fateev:2005,Fateev:2007b,Fateev:2008}. 
However, perhaps the best approach is to show that the integral of the absolute square of the instanton partition function in the quiver gauge theory satisfies the same differential equation as the corresponding Toda theory  correlation function.  As the prepotential is known to satisfy Picard-Fuchs type equations such an approach does not seem unreasonable. 

Let us also mention that there exists a conjectured relation in the Liouville theory between a certain four-point correlation function on the sphere and a one-point function on the torus (see (3.30) in \cite{Fateev:2009}). It may be possible to check this proposal using the gauge theory approach.

It would also be nice to find more examples of relations between 2$d$ CFTs and 4$d$ quiver gauge theories. There is a Toda theory associated with any Lie algebra. Are these related to quiver gauge theories? 

The most basic question is what the underlying reason for the Toda/quiver connection is.  Hopefully it will be possible to use the quiver gauge theories to learn more about the Toda theories and vice versa.

\section*{Acknowledgements}
The main part of this work was done while visiting Instituto Superior T\'ecnico, Lisbon whose hospitality is appreciated.
This work was supported by a grant from the Swedish Science Council.

\appendix

\setcounter{equation}{0} 
\section{Appendix}
In this appendix we collect some facts about instanton counting and $\cW$ algebras.
  
\subsection{Nekrasov Instanton counting}\label{appi}
The instanton counting method was developed by Nekrasov \cite{Nekrasov:2002} (further details can be found in~\cite{Bruzzo:2002,Nekrasov:2003}; see also \cite{Shadchin:2005a}). Below we focus on theories with a simple gauge group. More general quiver theories can also be treated in a similar manner. In particular, bifundamental matter has been treated in \cite{Fucito:2004,Alday:2009}.

The instanton partition function can be written
\be
Z_{\mathrm{inst}} = \sum_k q^k Z_{k}\,,
\ee
where for $\SU(N)$ 
\be \label{zint}
Z_{k} =  \frac{1}{k!} \int \prod_{i=1}^k \frac{\D \varphi_i}{2\pi i}\,, z_k(a,\varphi,\mu;\ep)\,.
\ee
Here $z_k$ depends on the field content of the model. The integrals in (\ref{zint}) can in many cases be performed explicitly leading to closed expressions for the $Z_k$. We will not give the details here. Instead we only give one example: the one instanton contribution in the $\SU(N)$ theory with $2N$ fundamentals can be written 
\be \label{suninun}
Z_1 = \frac{1}{\ep_1\ep_2} \sum_{i=1}^{N} \frac{M(\ha_i)}{\prod_{j\neq i} (\ha_i-\ha_j)(\ha_i-\ha_j+\ep)}\,,
\ee
where $\ep \equiv \ep_1+\ep_2$, $\sum_{i=1}^N \ha_i =0$ and 
\be
M(x) = \prod_{i=1}^N (x-\mu_i)(x+\bar{\mu}_i-\ep)\,.
\ee
It is convenient to write $\ha = \sum_{i=1}^{N-1} a_i \,e_i$ where $e_i$ are the simple roots of the $A_{N-1}$ Lie algebra. In the particular cases of $\SU(2)$ and $\SU(3)$, this translates into $\ha = (a,-a)$ and $\ha=(a_1,-a_1+a_2,-a_2)$, respectively. In the case of $\SU(2)$ we also write for the masses:
\be
\mu_1 = m_1 + \ka_1 \,, \qquad \mu_2 = m_2 + \ka_2 \,, \qquad  \bar{\mu}_1 = m_1-\ka_1   \,, \qquad  \bar{\mu}_2 = m_2-\ka_2  \,. 
\ee

In addition to the instanton contribution there is also a perturbative (1-loop) piece, $Z^{\rm pert}$, in the full partition function. $Z^{\rm pert}$ is a product of various factors. For $\SU(N)$ it is obtained from the building blocks (note that there is some freedom to redefine the $\mu_i$ and $\bar{\mu}_i$ by shifts):
\bea \label{zpert}
z^{\rm 1-loop}_{\rm gauge}(\ha) &=& \prod_{i<j} \exp[- \ga_{\ep_1,\ep_2}(\ha_i-\ha_j-\ep_1) -  \ga_{\ep_1,\ep_2}(\ha_i-\ha_j-\ep_2) ]\,, \non \\
z^{\rm 1-loop}_{\rm fund}(\ha,\mu) &=& \prod_{i} \exp[\ga_{\ep_1,\ep_2}(\ha_i-\mu) ] \,,\non\\
z^{\rm 1-loop}_{\rm antifund}(\ha,\bar{\mu}) &=& \prod_{i} \exp[\ga_{\ep_1,\ep_2}(\ha_i+\bar{\mu}-\ep) ]\,, \\
z^{\rm 1-loop}_{\rm adjoint}(\ha,\mu) &=& \prod_{i<j} \exp[\ga_{\ep_1,\ep_2}(\ha_i-\ha_j-\mu-\ep) ]\,. \non
\eea
Here the function $\ga_{\ep_1,\ep_2}(x)$ (not to be confused with $\ga(x)$ defined in (\ref{ga})) is related to $\Ga_2(x|\ep_1,\ep_2)(x)$ defined in (\ref{Gamma1}), (\ref{Gamma2}) as 
\be \label{gag}
\ga_{\ep_1,\ep_2}(x) = \log \Ga_2(x|\ep_1,\ep_2)\,.
\ee

In the case when the gauge group is $\Sp(2N)$ the $Z_k$'s are determined by
\be \label{spzint}
Z_{k} =   \frac{(-1)^k}{2^n n!} \int \prod_{i=1}^n \frac{\D \varphi_i}{2\pi i} \, z_n (a,\varphi,m;\ep)\,,
\ee
where $n =\lfloor  \frac{k-1}{2} \rfloor $ and  $\lfloor \cdot \rfloor $ denotes the integer part. We have\footnote{These expressions differ slightly from the ones in~\cite{Marino:2004} where the case $\ep=0$ was the main focus. The prescription used here arises from the method in \cite{Nekrasov:2004}} 
\be
z_k^{\mathrm{fund}}(a,\vphi,m;\ep) = (m-\ep/2)^{k-2n} \prod_{i=1}^n ( (m-(\ep/2))^2 - \vphi_i^2)\,,
\ee
and 
\bea
&&z_k^{\mathrm{gauge}}(a,\vphi;\ep) = \frac{\ep^n}{\ep_1^n\ep_2^n} {\left[ \frac{1}{2 \ep_1\ep_2 \prod_{l=1}^N ((\ep/2)^2 - a_l^2)} \prod_{i=1}^n \frac{\vphi_i^2(\vphi_i^2 - \ep^2)}{(\vphi_i^2 - \ep_1^2)(\vphi_i^2 -\ep_2^2)}\right]}^{k-2n} \non \\ && \quad \times \prod_{i=1}^n\frac{1}{P(\vphi_i - \ep/2)P(\vphi_i + \ep/2)(4\vphi_i^2 - \ep_1^2)(4\vphi_i^2 - \ep_2^2)} \\ && \quad \times \prod_{i<j} 
 \frac{(\vphi_i-\vphi_j)^2((\vphi_i-\vphi_j)^2-\ep^2) (\vphi_i+\vphi_j)^2((\vphi_i+\vphi_j)^2-\ep^2) }{ ((\vphi_i-\vphi_j)^2-\ep_1^2) ((\vphi_i-\vphi_j)^2-\ep_2^2) ((\vphi_i+\vphi_j)^2-\ep_1^2) ((\vphi_i+\vphi_j)^2-\ep_2^2) }  \non \,.
\eea
In the $\Sp(2N)$ case there is no restriction on the  $a_i$ and the index takes the values $i=1,\ldots,N$. It is not known how to write closed expressions for the above integrals for arbitrary instanton numbers (see \cite{Marino:2004} for a discussion), but at low orders in the instanton expansion the integrals can be explicitly performed.  

In the  $\Sp(2)$ theory the leading terms are  (using a convenient redefinition of the masses)
\be  \label{su2insp2} 
Z_{\rm inst} = 1 - q \frac{(m_1^2-\ka_1^2)(m^2_2-\ka_2^2))}{(4a^2 -\ep^2)}  +\cO(q^2)
\ee
Since $\SU(2)$ can also be viewed as $\Sp(2)$ this expression is an alternative to the one obtained viewing $\SU(2)$ as a subgroup of $\U(2)$. 

\subsection{$\cW$ algebra chiral blocks} \label{appw}

Here we illustrate how one calculates chiral blocks in a $\cW$ algebra, using the $\cW_3$ algebra as an example.. 
At level 1 a convenient basis of descendants is
\be
|\psi_1\rb =  L_{-1}|\al\rb\,, \qquad |\psi_2\rb= (W_{-1} - \frac{3w_\al}{2\De_\al} L_{-1})|\al\rb\,.
\ee
The $2\times2$ Gram matrix $\lb \psi_i|\psi_j\rb$ then becomes 
\be
\left( \ba{cc} 2 \De_\al   & 0 \\[5pt] 0  & \De_\al\left[\frac{32}{22+5c}(\De_\al +\frac{1}{5}) - \frac{1}{5}\right] - \frac{9}{2}\frac{w_\al^2}{\De_\al}  \ea \right),
\ee
and the chiral block of the four-point function in the $s$-channel becomes
\bea \label{block}
&&\frac{(\De_{\al_2} +\De_{\al} -\De_{\al_1})(\De_{\al_3} +\De_{\al} -\De_{\al_4} ) }{ 2 \De_{\al} } \non  \\
&+& \left( \frac{w_{\al_3}}{2} - \frac{w_{\al}}{2} - w_{\al_4}     +\frac{3}{2}\frac{\De_{\al}}{\De_{\al_3}}w_{\al_3} - \frac{3}{2}\frac{\De_{\al_4}}{\De_{\al_3}}w_{\al_3} - \frac{3}{2}\frac{\De_{\al_3}}{\De_{\al}}w_{\al}  + \frac{3}{2}\frac{\De_{\al_4}}{\De_{\al}}w_{\al}  \right) \non \\ &&\!\!\!\!\!\times 
\left( \frac{w_{\al_2}}{2} - \frac{w_{\al}}{2} - w_{\al_1}     +\frac{3}{2}\frac{\De_{\al}}{\De_{\al_2}}w_{\al_2} - \frac{3}{2}\frac{\De_{\al_1}}{\De_{\al_2}}w_{\al_2} - \frac{3}{2}\frac{\De_{\al_2}}{\De_{\al}}w_{\al}  + \frac{3}{2}\frac{\De_{\al_1}}{\De_{\al}}w_{\al}  \right)  \non \\
&&\!\!\!\!\!\times \left(\De_\al\left[\frac{32}{22+5c}(\De_\al +\frac{1}{5}) - \frac{1}{5}\right] - \frac{9}{2}\frac{w_\al^2}{\De_\al} \right)^{-1}.
\eea
Note that the first term is simply the standard Virasoro chiral block at this level. At higher levels the expressions quickly become very involved.

\begingroup\raggedright\endgroup

\end{document}